# Field-Induced Partial Disorder in a Shastry-Sutherland Lattice

Madalynn Marshall[1], Brianna R. Billingsley[2], Xiaojian Bai[1,4], Qianli Ma[1], Tai Kong[2,3], Huibo Cao[1*]

[1]Neutron Scattering Division, Oak Ridge National Laboratory, Oak Ridge, Tennessee 37831, USA

[2]Department of Physics, University of Arizona, Tucson, Arizona, 85721

[3]Department of Chemistry and Biochemistry, University of Arizona, Tucson, Arizona, 85721

[4]Department of Physics and Astronomy, University of Louisiana, Baton Rouge, Louisiana, 70803

*email: caoh@ornl.gov

**Abstract**

A 2-Q antiferromagnetic order of the ferromagnetic dimers was found below $T_N$ = 2.9 K in the Shastry-Sutherland lattice BaNd$_2$ZnS$_5$ by single crystal neutron diffraction. The magnetic order can be understood by the orthogonal arrangement of local Ising Nd spins, identified by polarized neutrons. A field was applied along [1 -1 0] to probe the observed metamagnetic transition in the magnetization measurement. The field decouples two magnetic sublattices corresponding to the propagation vectors $\mathbf{q_1}$= (½, ½, 0) and $\mathbf{q_2}$= (-½, ½, 0), respectively. Each sublattice shows a "stripe" order with a Néel-type arrangement in each single layer. The "stripe" order with $\mathbf{q_1}$ remains nearly intact up to 6 T, while the other one with $\mathbf{q_2}$ is suppressed at a critical field $H_c$ ~1.7 T, indicating a partial disorder. The $H_c$ varies with temperature and is manifested in the $H$-$T$ phase diagram constructed by measuring the magnetization in BaNd$_2$ZnS$_5$.

**Introduction**

Exotic non-trivial magnetic behavior has emerged in magnetic systems with geometrical frustration. The kagome and triangular lattices are common examples of two-dimensional (2D) frustrated lattices.[1,2] More elusive is the 2D orthogonal dimer lattice famously realized in the material $SrCu_2(BO_3)_2$,[3,4] which has been discovered to host a quantum spin liquid phase.[5–8] This lattice can be described by the Shastry-Sutherland (SS) model which consists of a 2D orthogonal arrangement of the spin dimers where the ratio between the two magnetic interactions, $\delta = J/J'$ with $J$ and $J'$ as the antiferromagnetic inter- and intra-dimer interactions, respectively, is critical for controlling magnetic states in the Shastry-Sutherland lattice (SSL).[9] The $\delta$ ratio of 0.675 and 0.765 separate the dimer singlet, plaquette singlet and Néel phase, respectively, in the phase diagram of the SS model.[10,11]

Resultantly, rich magnetic phase diagrams have been constructed for SSL materials from the field-induced evolution of the magnetic order. One common feature are fractionalized magnetization plateaus, which likely originate from the transition of a dimer singlet ground state to the formation of superstructures of the field-induced triplet dimers such as in $SrCu_2(BO_3)_2$.[10,12–21] The SSL is also found in families such as the rare earth tetraborides $RB_4$ (R=rare earth)[22–24], $BaR_2TO_5$ (R=rare earth, T=transition metal)[25–28], $R_2T_2In$ (R=rare earth, T=transition metal)[29–31] and $RE_2Pt_2Pb$ (R=rare earth)[32]. The magnetic ordered states can vary significantly from the insulator $SrCu_2(BO_3)_2$ which possesses a Heisenberg type exchange interaction to the metallic $RB_4$ family where a Ruderman-Kittel Kasuya-Yosida (RKKY) type interaction is observed between the moments giving it a long-rang order and possessing Ising-like moments oriented perpendicular to the SSL planes.[33–35] However, SSL materials such as $Yb_2Pt_2Pb$[32,36–39], have been found to exhibit field-induced metamagnetic transitions associated with partially disordered states, where at low temperatures a Luttinger liquid state has been realized.

Consequently, only very few SSL materials were reported to possess ferromagnetic dimers, including the insulator $BaNd_2ZnO_5$[25] and metallic $TmB_4$[23,34,40]. Considering the newly explored

BaR$_2$TO$_5$ family, although it is chemically diverse, the formation of the SSL will only occur for the lighter rare earth elements and successful single crystal growth has not been reported. The sulfide counterpart, BaR$_2$TS$_5$, which crystallizes into the same space group (*I*4/*mcm*) as the SSL BaR$_2$TO$_5$ materials, remains largely unexplored and recently large single crystals of BaNd$_2$ZnS$_5$[41,42] have been synthesized exhibiting a $T_N$ = 2.9 K. The dimers of the SSL in BaNd$_2$ZnS$_5$ are formed by the Nd atoms having inter-dimer lengths of 4.151 Å and intra-dimer lengths of 3.596 Å. A metamagnetic transition is observed in the *M*(*H*) with **H** along [1 -1 0], indicating this SSL material an excellent candidate for studying the intricate dimer physics from the field-induced magnetic phase evolution.

In this work, we report a 2-Q magnetic order in the SSL BaNd$_2$ZnS$_5$, determined by single crystal neutron diffraction. We utilize polarized neutrons to provide insight into the local magnetic anisotropy of the Nd spins and reveal the origin of the 2-Q magnetic order. The resulting field-induced evolution of the magnetic phases were characterized by magnetization measurements with the critical input from single crystal neutron diffraction. A partially disordered dimer liquid state was found and a "spin-flip-or-flop" mechanism was proposed to describe the dimer liquid state.

**Results and Discussions**

*Zero field Magnetic Structure*

BaNd$_2$ZnS$_2$ exhibits 2D SSL layers of magnetic Nd$^{3+}$ (*J* = 9/2) atoms separated by layers of Ba and Zn atoms and coordinated by S atoms. Nd$^{3+}$ has a Kramers doublet ground state and behaves as a pseudospin 1/2 [43]. Figure 1*a*, describes the Nd SSL where *J'* represents the intradimer interaction (nearest-neighbor) and *J* the interdimer interaction (next-nearest-neighbor), a typical

SS interaction model. To understand the magnetic anisotropy of the $Nd^{3+}$ spins in $BaNd_2ZnS_5$, we measured the local magnetic susceptibility tensor of the $Nd^{3+}$ spins by polarized neutron diffraction.[44,45] The local symmetry of the Nd atomic site, the 8h site of space group $I4/mcm$, implies the principal axes of the ellipsoid are along the [1 1 0], [1 -1 0] and [0 0 1] directions. The bulk magnetic measurements have revealed the magnetic moments are easy in-plane.[42] Therefore, only one-field-direction along [1 -1 0], was selected to detect the in-plane magnetic anisotropy of the $Nd^{3+}$ spins. By measuring reflections in spin-up and spin-down neutron channels, we obtained 17 good-quality flipping ratios to refine two free susceptibility tensor parameters in-plane. A suitable fitting of the flipping ratios could be reached, as shown by the Figure 1b plot of the experimental versus calculated flipping ratios, using the software CrysPy. The in-plane principal axes of the $Nd^{3+}$ magnetization ellipsoids, Figure 1c, were found to be orthogonal to the dimer bond consistent with an Ising-spin nature with lengths $\chi_{//} = 0.183(22)$ $\mu_B$/T and $\chi_\perp = 0.033(22)$ $\mu_B$/T. Similar to $Yb_2Pt_2Pb$[32] the magnetic moments are found within the plane of the SSL and orthogonally arranged between two magnetic sublattices that satisfy the Ising behavior, contrary to that in $SrCu_2(BO_3)_2$[14] and $TmB_4$[23]. Instead of the well-known SSL interaction model, the resulting formation seemingly favors an effective square lattice magnetic model where $J'$ and $J''$ are the potential interaction paths, as indicated by the orange line (i.e. the dimer bond) and the dashed black line, respectively, in Figure 1f.[32] The symmetric terms of interaction $J$ produce zero energy contribution due to the orthogonality of the spin arrangement between the neighboring orthogonal dimer bonds (shown in Figure 1a as a standard SS spin model but ignored in Figure 1f due to the zero-energy contribution from $J$ for the Ising spins). While the antisymmetric terms of $J$ in the spin Hamiltonian (see the full description in the SI) are likely weak as well, this will be shown by the field measurements presented later. Future inelastic neutron scattering measurements are needed to further confirm the interaction speculation here and interpret the spin dynamics in BaNd$_2$ZnS$_5$.

From the temperature-dependence of the magnetic scattering at (½ ½ 2), Figure 1*d*, the magnetic order appears at ~ 3 K, consistent with the reported $T_N$ = 2.9 K from the magnetic susceptibility measurements.[42] The solid red line in Figure 1*d*, corresponds to the power law fitting of the intensity, $I \sim (T_N - T/T_N)^{2\beta}$, with a $T_N$ reasonably fixed at 2.95 K, and a β ~ 0.08(1) which is smaller than the expected β = 1/8 for a 2D Ising system[32] and could be a result of the nature of the spin dimer lattice. As a product of the single crystal polarized neutron diffraction results, a suitable magnetic structure model could be immediately determined since the fit could be appropriately constrained with moments perpendicular to the dimer bonds. A 2-**Q** AFM model consisting of two magnetic sublattices indexed by the propagation vectors **q₁**= (½, ½, 0) and **q₂**= (-½, ½, 0), resembling the AFM 2-**Q** structure for BaNd$_2$ZnO$_5$[25] but with a different magnetic symmetry, best fit the zero-field data. The magnetic symmetry *P_C*4/*nnc* (#126.385) was then determined using the k-SUBGROUPSMAG program from the Bilbao Crystallographic server. Based on the body centered symmetry, the SSL layers are separated by a centering translation resulting in two inequivalent propagation vectors that each connects to one magnetic sublattice with a "stripe" order when viewing two layers together. Therefore, a FM interlayer interaction, $J_z$, also needs to be considered, which is likely as weak as *J″* due to the larger atomic distance between interacting spins, however, necessary to stabilize the magnetic order at zero field, distinct from the magnetic order reported in BaNd$_2$ZnO$_5$[25]. Each SSL layer individually exhibits a Néel phase arrangement where the potential interaction paths consist of a dominant FM *J′* and a weak AFM *J″*. The refined Nd magnetic moment was determined to be 2.6(1) μ$_B$. Figure 1*e* shows a plot of the calculated structure factor square ($F^2_{calc}$) versus the observed one ($F^2_{obs}$) and the magnetic structure can be seen in Figure 1*f* where the orange and blue atoms represent the two different sublattices and the overlapping layers along the *c* axis are indicated by the light and dark color shades.

*Field-Induced Phase Evolution*

The magnetization curves below $T_N$ of BaNd$_2$ZnS$_5$ show kinks around 1.7 T for fields along the [1 -1 0] direction, as shown by Figure 2*a*, indicating a metamagnetic transition. To investigate this transition, single crystal neutron diffraction measurements were performed with an applied magnetic field of 2 and 6 T parallel to the [1 -1 0] direction. As the field-induced transition emerges around 1.7 T, at 1.4 K, the field-dependent magnetic scattering at (1.5 1.5 1) disappears at 1.7 T (see Figure 2*c*) signifying the stripe phase with $q_2 = (-½, ½, 0)$ is no longer present. Therefore, the kink shown in the magnetization measurement is a signature of the magnetic order-disorder transition in the magnetic sublattice with spins parallel to the field and the corresponding field can be viewed as the critical field $H_C$ for this transition. For the other magnetic sublattice, the magnetic peak (½ ½ 4) gradually decreases with the field increasing but the majority of the magnetic peak signal is maintained up to 6 T (see Figure 2*c* inset). By analyzing the neutron diffraction data collected at 2 and 6 T at 1.4 K, the refined magnetic moments for the stripe phase of the $q_1$ magnetic sublattice with spins along [1 1 0], perpendicular to the field direction, were determined to be 2.8(1) and 2.6(1) $\mu_B$, respectively.

At 2 and 6 T the observed magnetic reflections from the diffraction patterns could be all indexed by $q_1 = (½, ½, 0)$, while no peaks could be indexed with $q_2 = (-½, ½, 0)$ when considering the body centering translation symmetry. Note, the body-centered unit cell is not a primitive cell and so $q_1$ and $q_2$ are not equivalent. From the k-SUBGROUPSMAG program, the low symmetry space group $P1$ (#1.1) was initially selected to test the potential magnetic models under field. The resulting refinement reveals a partially disordered state of ferromagnetic dimers at 2 T in one magnetic sublattice (Figure 2*e*) and while the AFM order in the other magnetic sublattice with moments along [1 1 0] ($\mathbf{q_1}$ magnetic sublattice) survives (Figure 2*e* and 2*f*). The results also indicate the two magnetic sublattices are interaction-decoupled and can be separated under an applied field, which confirms that the interaction $J$ between the two magnetic sublattices is weak, i.e., no strong antisymmetric exchange interactions between the orthogonally arranged neighboring Nd spin dimers. This separation can also be observed as two $T_N$'s connected to the two sublattices under field, see Supplementary Figure S1&S2. Considering the magnetic interaction distance and the localized *f*-electron feature for the rare-earth spins, both $J''$ and $J_z$ are much weaker compared to the intra-dimer interaction $J'$. Therefore, the SSL in BaNd$_2$ZnS$_5$ can be viewed as two decoupled square lattices of ferromagnetic dimers that are loosely 3-dimensionally connected. While the magnetic phase transitions induced by the field up to 6 T along [1 -1 0] at 1.4 K is likely only within the $\mathbf{q_2}$ magnetic sublattice with Nd$^{3+}$ moments along [1 -1 0]. Therefore, the following discussions will be focused on only the magnetic dimer square sublattice with spins parallel to the field [1 -1 0] (the square lattice of FM spin dimers is shown in Figure 2e and 2*f* as the dark orange line for the magnetic sublattice with spins along [1 -1 0]).

Neutron diffraction revealed that the $\mathbf{q_2}$ stripe order is fully suppressed by the field at the critical field $H_C \sim 1.7$ T. Field-induced magnetic signal on top of the nuclear Bragg peaks were refined as uniformly aligned moments of 1.2(2) $\mu_B$ at 2 T for the magnetic sublattice with the spin Ising axis along [1 -1 0] // **H**, i.e., 0.6(1) $\mu_B$ per Nd$^{3+}$ if averaging it for the whole magnetic lattice, consistent with the increased magnetization in the bulk measurement. If we consider the model of the square lattice of ferromagnetic dimers as described above, each ferromagnetic dimer includes two parallel aligned spin-half moments and so makes a spin-1 dimer with $S = 1$ as the ground state. When a field is applied, two kinds of dimer spin transitions among three magnetic components ($S_Z = -1, 0, +1$) can occur and cause the order-disorder transition and yield the average induced moment as seen by neutrons. One is a spin-flip transition from $S_Z = -1$ to $S_Z = +1$ and the other one is a spin-flop transition from $S_Z = -1$ to $S_Z = 0$, the quantum version of the well-known spin-flop transition in a weak-magnetic-anisotropic AFM system.[46,47] Both spin component transitions are illustrated in Figure 3b. The flipped or flopped spin dimers are disordered in the lattice space because no additional superlattice ordering peaks were observed. Therefore, we refer to this partial disorder as a liquid-like state, i.e., dimer liquid. More dynamic measurements are needed to further characterize the nature of the dimer liquid state in contrast to the reported Luttinger liquid state in the SSL material $Yb_2Pt_2Pb$[48], the quantum spin liquid phase in the Kitaev honeycomb-lattice $RuCl_3$[49], and the magnetization plateau phases in the SSL material $SrCu_2(BO_3)_2$[10,12–21]. At 6 T the square magnetic sublattice with Nd spins along [1 -1 0] enters a field polarized state and has a refined magnetic moment of 2.8(1) $\mu_B$, the magnetic structure is shown in Figure 2f.

*Phase Diagram and Spin Dimer Liquid*

We can construct the field-temperature ($H$-$T$) phase diagram from the magnetization data upon varying temperature since neutron diffraction reveals that the kink observed in the magnetization indicates the disorder transition from the stripe order in the $q_2$ magnetic lattice. The critical field $H_C$ at each temperature can be better observed as a sharp peak in the plots of d$M$/d$H$ (see Figure 2b), where upon cooling, the $H_C$ shifts towards higher fields, $H_C$ ~1.5 T at 1.8 K. Above $H_C$, the transition to the field polarized state can be seen by the broad bump feature at higher field. A contour plot mapping the values of d$M$/d$H$ obtained from 1.8 – 3.8 K, is depicted in a $H$ versus $T$ ($H$-$T$) phase diagram describing the magnetic sublattice with spins parallel to the field direction

[1 -1 0], Figure 3a, based on the bulk magnetization (circles) and susceptibility (diamonds and triangles) measurements under the field along [1 -1 0] and neutron data (square). Phase I represents the 2-**Q** magnetically ordered stripe phase while phase II is the dimer liquid phase. The contour plot of d$M$/d$H$ clearly defines the regions for Phase I and II and furthermore the transition to the field polarized state as Phase III. According to the neutron diffraction measurement at 1.4 K, no additional magnetic order was observed when the stripe phase enters the spin dimer liquid state, Phase II, at 1.7 T, indicating the critical region is further narrowed towards a possible critical point.

A similar temperature dependence of the behavior of $H_C$ in Figure 3a has been observed in the phase diagrams of geometrically frustrated lattices exhibiting field-induced quantum criticality, such as $CoNb_2O_6$[50,51] and $RuCl_3$[49], constructed from heat capacity measurements under field. A possible explanation of such a *H-T* behavior in $BaNd_2ZnS_5$ can be explained by field-melting the stripe ordered phase through "spin-flip-or-flop" transitions. The *H-T* phase diagram could also suggest the possibility that quantum criticality may exist in $BaNd_2ZnS_5$ at the lowest temperature. Additional evidence is shown by the plot of d$M$/d$H$|*max* versus *T* in the Figure 2b inset that demonstrates the power law fitting ~ $T^{-n}$, where it has been constrained to n = 3.1 to assess the possibility of quantum critical fluctuations as determined for the antiferromagnet $CePtIn_4$[52]. Down to 1.8 K $BaNd_2ZnS_5$ resembles that of the linear higher temperature region in the power law fitting of $CePtIn_4$, suggesting a certain similarity of the two systems. To explore these scenarios and reveal the enigmatic magnetic ground state of $BaNd_2ZnS_5$, low temperature heat capacity, magnetization and inelastic neutron scattering measurements are required. In summary, single crystals of $BaNd_2ZnS_5$ from the $BaR_2ZnS_5$ family have been successfully synthesized and studied by magnetic bulk measurement and neutron diffraction under magnetic field. Ising magnetic anisotropy of the Nd spins is revealed by the local magnetic susceptibility method with polarized neutrons and their Ising directions are orthogonal to the dimer bonds. Such an arrangement of Ising spins implies symmetric exchange interactions cannot couple the orthogonally arranged dimers. The zero-field magnetic order is a 2-**Q** AFM order of FM dimers with Nd spins along their local Ising directions. Two magnetic sublattices with **q$_1$**= (½, ½, 0) and **q$_2$**= (-½, ½, 0) constitute the 2-**Q** structure and respond to the field along [1 -1 0] differently. The **q$_2$** magnetic order is suppressed at a critical field, responsible for the kink observed in the magnetization measurement. While the **q$_1$** magnetic sublattice stays mostly unchanged until 6 T. According to this information, we built the *H-T* phase diagram from the bulk magnetization and magnetic susceptibility. A critical region was manifested as a spin dimer liquid phase growing out between the stripe phase at lower field and the field polarized phase at upper field. "Spin-flip" and "spin-flop" mechanisms were proposed to explain the formation of the liquid state. Whether the dimer liquid phase could condense at an ultra-low temperature deserves further study. As BaNd$_2$ZnS$_5$ exhibits a unique 2-**Q** magnetic square lattice with weak interactions between orthogonal dimers, it can be an exciting candidate to exhibit unique high order symmetries and a potential host for exotic quantum phases. Future dynamic studies including inelastic neutron scattering can provide insight into magnetic interactions and emergent states in the BaNd$_2$ZnS$_5$ SSL and discover interesting dynamic properties of ferromagnetic dimers. An effort in synthesizing large, high quality single crystals is thus called for, as available high-quality single crystals make all these possible.

## Methods

Neutron Diffraction

To determine the magnetic order single crystal neutron diffraction experiments on the HB-3A DEMAND[53] at the High Flux Isotope Reactor at Oak Ridge National Laboratory. A 1.5 mm sample was measured with the two-axis mode down to 1.4 K using a cryomagnet and a wavelength of 1.542 Å from a bent Si-220 monochromator[54]. The measurement was performed with an applied magnetic field of 0, 2 and 6 T parallel to [1 -1 0]. The Bilbao Crystallography Server[55] was used for the magnetic symmetry analysis and Fullprof software[56] for the magnetic structure refinement. Polarized single crystal neutron diffraction measurements was likewise performed on the HB-3A DEMAND with a polarized neutron beam of 1.542 Å and the calibrated neutron polarization is 72%. The crystal was loaded in a closed-cycle refrigerator with a permanent magnet set providing the fixed field of 0.5 T along [1 -1 0]. We measured 17 flipping ratios at 5 K, above $T_N$. To analyze the resulting flipping ratios the CrysPy software[57] was used.

## Data Availability

All relevant data are available from the corresponding author upon reasonable request.

**Acknowledgements**

The research at Oak Ridge National Laboratory (ORNL) was supported by the U.S. Department of Energy (DOE), Office of Science, Office of Basic Energy Sciences, Early Career Research Program Award KC0402020, under Contract DE-AC05-00OR22725. This research used resources at the High Flux Isotope Reactor, a DOE Office of Science User Facility operated by ORNL. We acknowledge and thank Zachary Morgan for the assistance with the data reduction of the neutron diffraction experiments. We acknowledge and thank Cristian Batista for the helpful discussions. We thank Chenyang Jiang for the assistance with the experimental polarized neutron diffraction.

**Author Contributions**

H.C. cultivated the research; B.R.B and T.K. synthesized single-crystal samples and performed magnetization measurements; M.M. and H.C. performed neutron scattering experiments and analyze the data; X.B. assisted with polarized neutron data analysis; X.B. and Q.M. contributed to theoretical discussion; M.M. and H.C. wrote manuscript with comments from all authors.

**Competing Interests Statement**

The authors declare no competing interests.

This manuscript has been authored by UT-Battelle, LLC under Contract No. DE-AC05-00OR22725 with the U.S. Department of Energy. The United States Government retains and the publisher, by accepting the article for publication, acknowledges that the United States Government retains a non-exclusive, paidup, irrevocable, world-wide license to publish or reproduce the published form of this manuscript, or allow others to do so, for United States Government purposes. The Department of Energy will provide public access to these results of federally sponsored research in accordance with the DOE Public Access Plan(http://energy.gov/downloads/doepublic-access-plan).

**Additional Information**

Corresponding Author

Correspondence to Dr. H. Cao (caoh@ornl.gov).

**Figure Legends/Captions**

**Figure 1.** Zero-field magnetic structure refinement of $BaNd_2ZnS_5$. **a** SSL sublattice of Nd atoms in the *ab* plane. **b** Experimental versus calculated flipping ratio plot. **c** Local magnetic anisotropy of Nd dimers is showed by magnetic susceptibility tensors drew as ellipsoids in unit cell of $BaNd_2ZnS_5$. **d** Temperature-dependent order parameter of peak (½ ½ 2), red line is the empirical power law fitting, $I \sim (T_m - T/T_m)^{2\beta}$. **e** Experimental versus calculated structure factors at zero field. **f** 2-**Q** magnetic structure model for $BaNd_2ZnS_5$, the blue and orange represent the sublattices with $\mathbf{q_1}$= (½, ½, 0) and $\mathbf{q_2}$= (-½, ½, 0), respectively, and the light and dark color shades represent the different layers along the *c* axis. The $J'$ and $J''$ interaction paths correspond to the dimer bond and the dashed black line, respectively. The $J_z$ represents the nearest spin-spin interaction between layers.

**Figure 2.** Field induced phase evolution and magnetic structure refinement at 2 and 6 T. **a** Magnetization measurement with **H**//[1 -1 0] from 1.8 – 3.8 K. **b** Plots of d$M$/d$H$ measured at constant temperature from 1.8-3.8 K, data was smoothed for the derivation, with an inset of d$M$/d$H$|*max* versus temperature as derived from the maximum value of the sharp peak in d$M$/d$H$, the red line represents the power-law dependence ~ $T^{-n}$ (n=3.1). **c** Field-dependent order parameter from 0-2 T at (1.5 1.5 1). The inset for the field-dependent order parameter from 2-6 T at (½ ½ 4). **d** Experimental versus calculated structure factors at 2 T (red circles) and 6 T (blue triangles). **e** The refined magnetic structure of the partially disordered phase at 2 T and **f** the field polarized state at 6 T with the square lattice of FM spin dimers shown by the dashed dark orange lines.

**Figure 3.** A constructed phase diagram along with possible dynamical magnetic patterns for disordered state. **a** Field versus $T_N$ (*H-T*) phase diagram with **H**//[1 -1 0], the magnetic susceptibility, magnetization and neutron data are represented by the light blue diamonds and blue triangles for the two measurements which were performed, yellow circles and green squares, respectively. The *H-T* phase diagram overlays a contour plot mapping the values of d$M$/d$H$ obtained at constant temperatures, the values of d$M$/d$H$ are represented by the color scale. **b** A possible dynamical magnetic pattern at the $H_C$, the red-circle highlights the spin-flip transition from $S_Z = -1$ to $S_Z = +1$ and the black-circle highlights the spin-flop transition from $S_Z = -1$ to $S_Z = 0$, the orange circles symbolize no magnetic moment is present.

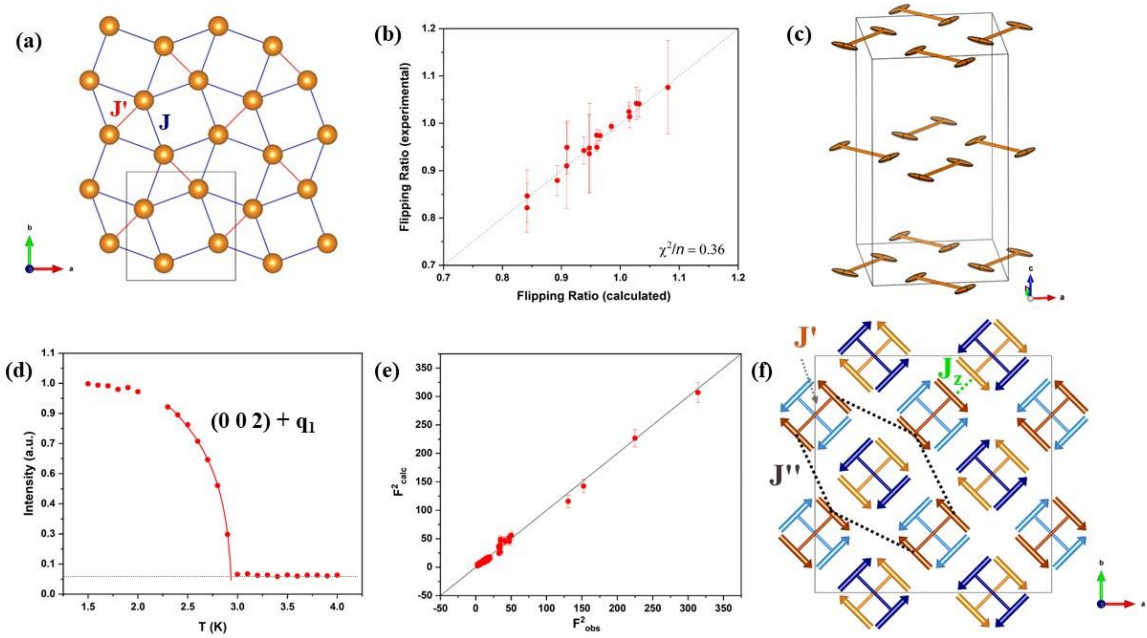

**Figure 1.**

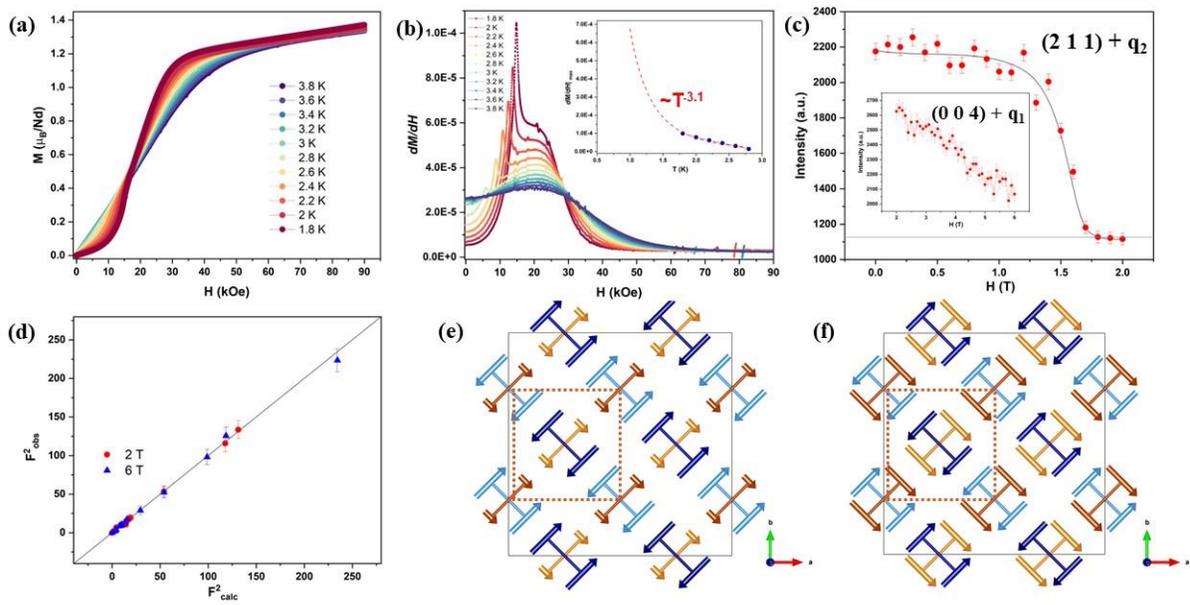

**Figure 2.**

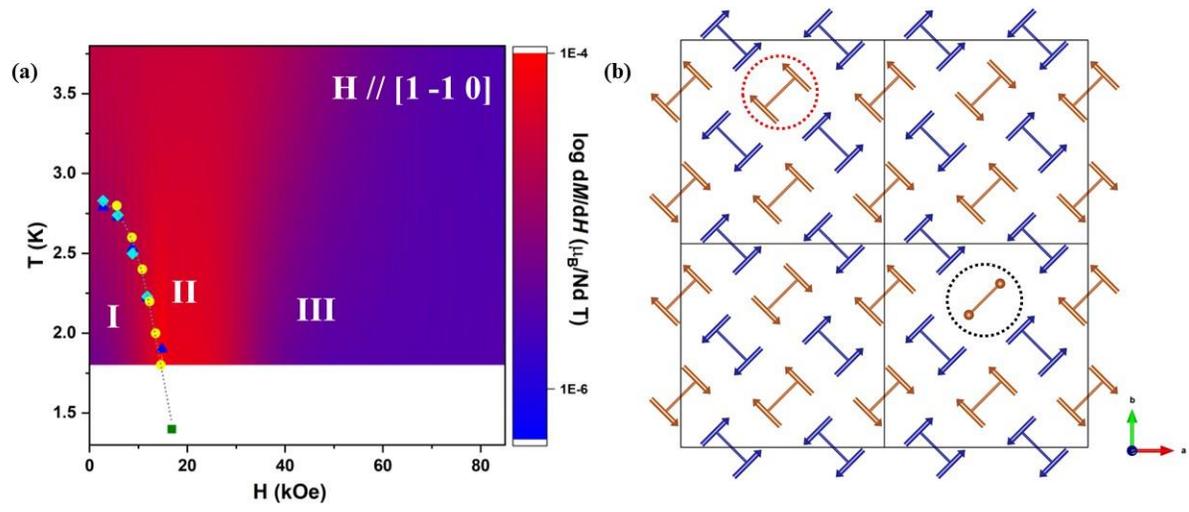

**Figure 3.**

# Supplementary Information

Field-Induced Partial Disorder in a Shastry-Sutherland Lattice

Madalynn Marshall[1], Brianna R. Billingsley[2], Xiaojian Bai[1,4], Qianli Ma[1], Tai Kong[2,3], Huibo Cao[1*]

[1]Neutron Scattering Division, Oak Ridge National Laboratory, Oak Ridge, Tennessee 37831, USA

[2]Department of Physics, University of Arizona, Tucson, Arizona, 85721

[3]Department of Chemistry and Biochemistry, University of Arizona, Tucson, Arizona, 85721

[4]Department of Physics and Astronomy, University of Louisiana, Baton Rouge, Louisiana, 70803

*email: caoh@ornl.gov

1. **Spin Hamiltonian**

By considering $J$ - the nearest orthogonal spin interactions, $J'$ - intradimer spin interactions, $J''$ - the 2$^{nd}$ nearest interdimer interactions, $J_z$ – the nearest interlayer spin interactions (see Figure 1$a$ and 1$f$ in the main manuscript), the spin Hamiltonian in the SSL $BaNd_2ZnS_5$ is expressed as,

$$H = \sum_{nn'} S_n \cdot J \cdot S_{n'} + \sum_{ii'} S_i \cdot J' \cdot S_{i'} + \sum_{jj'} S_j \cdot J'' \cdot S_{j'} + \sum_{kk'} S_k \cdot J_z \cdot S_{k'}$$

According to the bond symmetry analysis using Su(n)ny [1], the interaction tensors are listed as

$$J = \begin{bmatrix} X & Z+P & 0 \\ Z-P & X & 0 \\ 0 & 0 & Y \end{bmatrix}, J' = \begin{bmatrix} A & D & 0 \\ D & B & 0 \\ 0 & 0 & C \end{bmatrix}, J'' = \begin{bmatrix} E & H & 0 \\ H & F & 0 \\ 0 & 0 & G \end{bmatrix},$$

$$J_z = \begin{bmatrix} I & K+L & M-N \\ K-L & I & -M-N \\ M+N & -M+N & R \end{bmatrix}.$$

and 18 non-zero elements represent the allowed interactions.

As discussed in the "*Field-Induced Phase Evolution*" section in the main manuscript, the neutron diffraction data suggest the FM interaction $J'$ is dominant and the inter-dimer interaction $J$ is weak, which decouples two magnetic sublattices as shown by the field-dependence of magnetic orders. While $J''$ and $J_z$ interactions in each magnetic sublattice are required to stabilize the observed magnetic order although they are not expected to be strong due to the larger bond distances. This leaves the effective Hamiltonian to be $H = \sum_{ii'} S_i \cdot J' \cdot S_{i'} + \sum_{jj'} S_j \cdot J'' \cdot S_{j'} + \sum_{kk'} S_k \cdot J_z \cdot S_{k'}$.

Within the context of the main manuscript, we discussed the interactions using the "symmetric" exchange ($X,Y,Z$; $A,B,C,D$; $E,F,G,H$; $I,R,K,M$) and "antisymmetric" exchange ($P$, $L$, and $N$) in the matrix representation of $J$, $J'$, $J''$ and $J_z$ and evaluated the relative interaction strengths according to the diffraction data. To quantitatively determine the interaction matrix of the spin Hamiltonian, one needs to measure spin dynamics by inelastic neutron scattering, which requires larger crystals.

## 2. Field-dependent specific heat and $T_N$

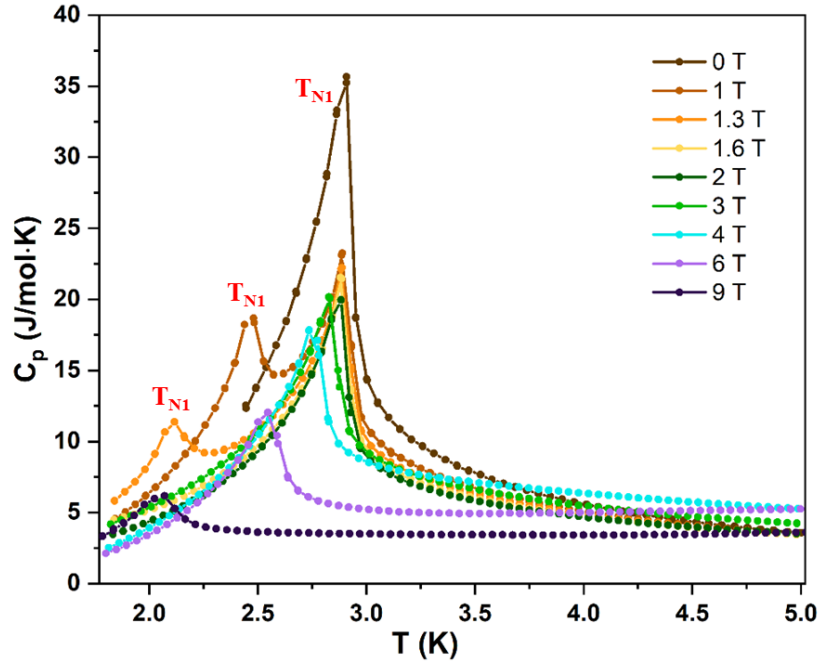

**Supplementary Figure S1.** Bulk temperature dependent specific heat measurements down to 1.8 K for a series of fields from 0 – 9 T. Measurements were performed on a Quantum Design physical property measurement system (Dynacool) using the two-tau relaxation method on a 2.7mg sample with **H** // (1 1 0). As indicated by peaks in each specific plot, one magnetic transition splits into two under field. The transition associated with the magnetic sublattice with spins parallel to the field [1 -1 0] is labeled as $T_{N1}$. $T_{N2}$ is for the other one with spins perpendicular to **H**.

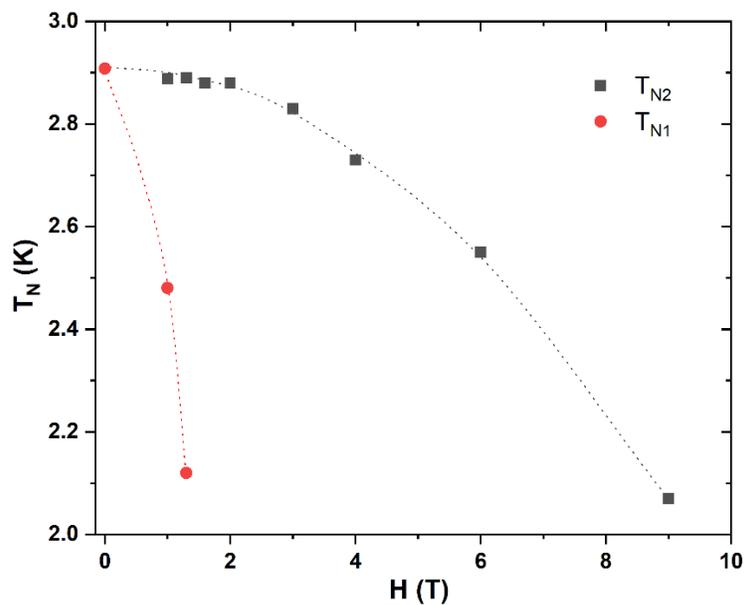

**Supplementary Figure S2.** Transition temperature ($T_N$) versus field **H** // (1 1 0). The data are extracted from the bulk temperature dependent specific heat measurements for a series of fields from 0 – 9 T as shown in Figure S1.

3. **Magnetization without hysteresis**

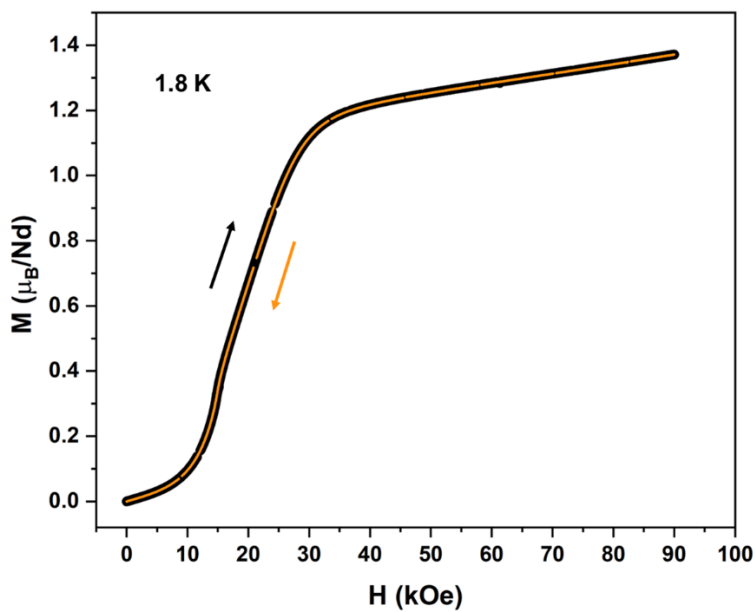

**Supplementary Figure S3.** Magnetization measurement shows no hysteresis observed at 1.8 K, the black and orange line represents the field wrapping up and down as indicated by the arrows.